\documentclass[12pt]{article}

\usepackage{latexsym}

\begin{document}

\title{The covariant graviton
propagator in de\ Sitter spacetime}

\author{Atsushi Higuchi$^1$ and Spyros S.\ Kouris$^2$\\ 
{\normalsize Department of Mathematics, University of York}\\ 
{\normalsize Heslington, York, YO10 5DD, United Kingdom}\\
{\normalsize $^1$Email: ah28@york.ac.uk}\\
{\normalsize $^2$Email: ssk101@york.ac.uk}}

\date{9 September, 2001}
\maketitle

\begin{abstract} 
We consider the covariant graviton propagator in de\ Sitter spacetime in a
gauge with two parameters, $\alpha$ and $\beta$, in the Euclidean approach.
We give an explicit form of the propagator with a particular choice of $\beta$
but with arbitrary value of $\alpha$.  We confirm that two-point functions
of local gauge-invariant quantities do not increase as the separation of the
two points becomes large.
\end{abstract}

\section{Introduction}

Quantum field theory in de Sitter spacetime (see, e.g., \cite{HawkEllis}) 
has been studied extensively
because of its relevance to inflationary 
cosmologies~\cite{Guth,Linde,Stein}.
The graviton two-point functions, which represent correlation of vacuum
fluctuation in the gravitational field, 
have been studied by many authors using various gauges in this spacetime
(see, e.g., \cite{Allen86:4}--\cite{AIT}).  
These two-point functions are known to
increase as the (coordinate)
distance between the two points becomes large.  Some authors have
suggested that this behaviour may manifest itself 
in the two-point functions of
gauge invariant quantities. 

However, a non-covariant two-point function which
does not increase as a function of two-point distance was obtained recently
in open de\ Sitter spacetime~\cite{HHT}.  
Subsequently, the present authors showed that the 
logarithmically increasing term of the non-covariant graviton two-point 
function in the spatially-flat coordinates found by 
Allen~\cite{Allen87:1} is a gauge artifact~\cite{HigKou}. 
It is desirable to extend these observations to 
covariant gauges, which are more suitable for computation. 
The present authors have started investigation in this direction and 
shown that the pure-trace part of a covariant two-point
function, whose growth with distance was suggested to be
physical~\cite{AllenTuryn,AM}, 
gives a pure-gauge contribution when combined with another
part~\cite{Higuchi87:1,HigKou2}.  

In this paper we go beyond that work and consider the
full two-point function in a covariant gauge with two parameters.
(This gauge was previously used in \cite{AIT}, but the explicit
form of the corresponding two-point function has not been written down.) 

We work in the Euclidean approach of Allen and Turyn~\cite{AllenTuryn}.
(Thus, what we compute is the Green function on the
4-sphere, which becomes
the Feynman propagator in de\ Sitter spacetime upon analytic continuation. 
It can be identified with the Wightman two-point 
function for two points which are not causally connected.)
These authors chose the gauge in which the flat-space limit is simplest and
found that the propagator increases as the separation between the
two points becomes large.
We generalize their work and find the propagator which depends on
two gauge parameters, $\alpha$ and $\beta$.
Then we write down the propagator explicitly
for a particular choice of $\beta$ with the value of $\alpha$ left arbitrary.
(Our propagator consists of three sectors: the transverse-traceless, vector
and scalar sectors.   
The transverse-traceless and vector sectors are the same as those 
in \cite{AllenTuryn}.  We need to generalize only the scalar sector.) 
Unfortunately, we cannot find any choice of gauge parameters that
eliminates the large-distance 
growth of the propagator. Nevertheless, we show that this
growth will not be reflected in the two-point function of a local 
gauge-invariant quantity, as expected from the results in non-covariant
gauges mentioned before.

The rest of the paper is organized as follows. In section 2 we discuss the
general structure of the Green function following Allen and Turyn.
We write down their results
for the transverse-traceless and vector sectors of the
propagator in section 3 for completeness, and present our result for the
scalar sector in section 4.  In the final section we summarize this paper 
and show that the two-point function of a local gauge-invariant quantity is
bounded as the separation of the two points becomes large.
We use natural units $\hbar = c = 1$ throughout this paper.

\section{The field equation and the Green function}

De\ Sitter spacetime is a contracting and
expanding 3-sphere with the following line element:
\begin{equation}
ds^2=-dt^2+\frac{1}{H^2}\cosh^2(Ht)
[d\chi^2+\sin^2\chi(d\theta^2+\sin\theta^2d\phi^2)]\,, \label{metric}
\end{equation}
where $H$ is the Hubble constant.
By introducing the variable $\tau \equiv \pi/2-iHt$, 
we obtain
\begin{equation}
ds^2=H^{-2}\left\{
d\tau^2+\sin^2\tau\,[d\chi^2+\sin^2\chi(d\theta^2+\sin\theta^2d\phi^2)]
\right\}\,,
\end{equation}
which is the line element of a four-dimensional sphere ($S^4$)
of radius $H^{-1}$.
We will compute the Green function of the linearized gravity on
this space, which becomes the Feynman propagator in de\ Sitter spacetime
upon analytic continuation.  (For this reason we will use the terms 
``Green function" and ``propagator" interchangeably.)
Then, we are automatically choosing 
the Euclidean~\cite{GibHawk}, or Bunch-Davies~\cite{BunchDavies},
vacuum.

We start from the Lagrangian density of pure gravity with positive cosmological
constant,
\begin{equation}
{\cal L}_{\rm full} = \sqrt{-\tilde{g}}\,(R -6H^2)\,, \label{fullL}
\end{equation} 
where $\tilde{g}_{ab}$ is the full metric and where $R$ is the corresponding
scalar curvature.
We write the metric as $\tilde{g}_{ab} = g_{ab} + h_{ab}$, 
where $g_{ab}$ is 
the background de\ Sitter metric given by (\ref{metric}), and expand
the Lagrangian (\ref{fullL}) to second order in $h_{ab}$.
The Lagrangian density for the linearized gravity thus obtained is 
written, after
dropping a total divergence, as
\begin {eqnarray} 
\mathcal{L}_{\rm inv} &=& \sqrt{-g}
\left[ \frac{1}{2}\nabla_{a}h^{ac}\nabla^{b}h_{bc}
-\frac{1}{4}\nabla_{a}h_{bc}\nabla^{a}h^{bc}
+\frac{1}{4}(\nabla^{a}h-2\nabla^{b} h^{a}_{\ b})\nabla_{a}h \right.
 \nonumber \\
&& \left. \ \ \ \ \ \ \ \ \ \ 
-\frac{1}{2}H^2\left(
h_{ab}h^{ab}+\frac{1}{2}h^2\right)\right] \label{Lagden}
\end{eqnarray}
with $h = h^{a}_{\ a}$.  Here 
the indices are raised and lowered by $g_{ab}$, and
the $\nabla_a$ denotes the background covariant derivative. 
This Lagrangian density is invariant
under the gauge transformation
\[
h_{ab}\to h_{ab}+\nabla_{a}\Lambda_{b}+\nabla_{b}\Lambda_{a}
\]
up to a total divergence.
One needs to break this gauge invariance
for canonical quantization. 
For this purpose we add the following
gauge-fixing term in the Lagrangian density:
\begin{equation}
\mathcal{L}_{\rm gf}=-\frac{\sqrt{-g}}{2\alpha}
\left( \nabla_a h^{ab}-\frac{1+\beta}{\beta}\nabla^{b}h\right)
\left( \nabla^c h_{cb}-\frac{1+\beta}{\beta}\nabla_{b}h\right)\,.
\end{equation}
Then the Euler-Lagrange field equations derived from
${\cal L}_{\rm inv}+{\cal L}_{\rm gf}$ are 
\begin{eqnarray}
 {L_{ab}}^{cd}h_{cd}
&\equiv &\frac{1}{2}\Box h_{ab}-\left(\frac{1}{2}-\frac{1}{2\alpha}\right)
\left(\nabla_{a}\nabla_{c} h^{c}_{\ b}
+ \nabla_{b}\nabla_{c} h^{c}_{\ a}\right) \nonumber \\
&& +\left[\frac{1}{2}
-\frac{\beta+1}{\alpha\beta}\right]
 \nabla_a\nabla_b h+\left[\frac{(\beta+1)^2}{\alpha\beta^2}
-\frac{1}{2}\right]g_{ab}\Box h
  \nonumber \\
&& + \frac{1}{2}g_{ab}\left(1-\frac{2(1+\beta)}{\alpha\beta}\right)
\nabla_{c}\nabla_{d} h^{cd}
-H^2\left(h_{ab}+\frac{1}{2}g_{ab}h\right) = 0\,.
\label{Eulag1}
\end{eqnarray} 

To find the Green function of the operator $-{L_{ab}}^{cd}$ on $S^4$
we need the symmetric tensor eigenfunctions
of the Laplace-Beltrami operator $\Box = \nabla_a \nabla^a$ described in
\cite{AllenTuryn}.  It is convenient to define the
inner product of any two symmetric tensors $h_{ab}^{(1)}$ and 
$h_{ab}^{(2)}$ on $S^4$ as
\begin{equation}
(h^{(1)},h^{(2)})_T = \int_{S^4} dx\, \overline{h_{ab}^{(1)}}h^{(2)ab}\,,
\end{equation}
where $dx$ is the volume element.
The inner product of scalar functions, $(\cdot,\cdot)_S$, and
that of vector functions, $(\cdot,\cdot)_V$, are defined in a similar manner.

The scalar eigenfunctions $\phi^{(n,i)}$
satisfy
\begin{equation}
\Box \phi^{(n,i)} = - \lambda_n\phi^{(n,i)}\,,
\end{equation}
where $\lambda_n = n(n+3)H^2$, $n=0,1,2,\ldots$, and where the index
$i$ distinguishes the modes with the same eigenvalue $\lambda_n$. (It is
well known that the spectrum of the operator $\Box$ on the $N$-sphere of radius
$H^{-1}$ is given by 
$-n(n+N-1)H^2$, where $n$ is a non-negative integer.
See, e.g., \cite{ChodosM}.)  We
impose the normalization condition 
$(\phi^{(n,i)},\phi^{(m,j)})_S = \delta^{nm}\delta^{ij}$.
The pure-trace 
eigenfunctions of the operator $\Box$ on the symmetric tensors are 
\begin{equation}
\chi_{ab}^{(n,i)} = \frac{1}{2}g_{ab}\phi^{(n,i)}\,. \label{defchi}
\end{equation}
The traceless eigenfunctions that can be expressed in terms of scalars are
\begin{equation}
W_{ab}^{(n,i)} = \frac{2(\nabla_a \nabla_b-\frac{1}{4}g_{ab}\Box)
\phi^{(n,i)}}
{\sqrt{3\lambda_n(\lambda_n-4H^2)}}\,,\ \
 (n\geq 2)\,. \label{defW}
\end{equation}  
The modes with $n\leq 1$ are missing because the differential operator
$\nabla_a\nabla_b - \frac{1}{4}g_{ab}\Box$ annihilates $\phi^{(n,i)}$ with
$n\leq 1$. These tensor 
eigenfunctions satisfy $(\chi^{(n,i)},W^{(m,j)})_T = 0$ and
$(\chi^{(n,i)},\chi^{(m,j)})_T=(W^{(n,i)},W^{(m,j)})_T=\delta^{nm}\delta^{ij}$.
The divergence-free vector eigenfunctions $\xi^{(n,i)}$ satisfy~\cite{ChodosM} 
\begin{equation}
\Box \xi^{(n,i)} = - (\lambda_n - H^2)\xi^{(n,i)}\,,\ \ (n\geq 1).
\end{equation}
We impose $(\xi^{(n,i)},\xi^{(m,j)})_V = \delta^{nm}\delta^{ij}$. 
The symmetric tensor eigenfunctions that are expressed in
terms of these vectors are
\begin{equation}
V_{ab}^{(n,i)} = \frac{\nabla_a\xi_b^{(n,i)} + \nabla_b\xi_a^{(n,i)}}
{\sqrt{2(\lambda_n - 4H^2)}}\,,\ \ (n\geq 2)\,. 
\end{equation}
The modes with $n=1$ are missing because the vectors $\xi^{(1,i)}_a$ satisfy
$\nabla_a \xi^{(1,i)}_b + \nabla_b \xi^{(1,i)}_a=0$.
These tensor
eigenfunctions are orthogonal to the modes discussed before with respect
to the inner product $(\cdot,\cdot)_T$ and satisfy 
$(V^{(n,i)}, V^{(m,j)})_T = \delta^{nm}\delta^{ij}$.
The traceless and divergence-free (transverse-traceless) modes 
$h^{(TT,n,i)}_{ab}$ satisfy~\cite{ChodosM}
\begin{equation}
\Box h^{(TT,i,n)}_{ab} = - (\lambda_n - 2H^2)
h^{(TT,i,n)}_{ab}\,,\ \ (n\geq 2),
\end{equation}
and $(h^{(TT,n,i)}, h^{(TT,m,j)})_T = \delta^{nm}\delta^{ij}$.  These
modes are orthogonal to the modes $\chi^{(n,i)}_{ab}$, $W^{(n,i)}_{ab}$
and $V^{(n,i)}_{ab}$ with repspect to $(\cdot,\cdot)_T$.
Any symmetric tensor field on $S^4$ can be expanded 
in terms of these eigenfunctions.

Now, define the $\delta$-function $\delta_{aba'b'}(x,x')$ by
\begin{equation}
\int \delta_{aba'b'}(x,x')f^{a'b'}(x')dx'=f_{ab}(x)
\end{equation}
for any smooth symmetric tensor $f_{ab}$ on $S^4$.
Here, $dx'$ is the volume element.
Then, using the completeness of the modes described before, 
we can express the $\delta$-function as
\begin{equation}
\delta_{aba'b'}(x,x') = \delta^{(TT)}_{aba'b'}(x,x')
+ \delta^{(V)}_{aba'b'}(x,x') + \delta^{(S)}_{aba'b'}(x,x')\,,
\end{equation}
where
\begin{eqnarray}
\delta^{(TT)}_{aba'b'}(x,x') & = & 
\sum_{n=2}^{\infty}\sum_{i}
h^{(TT,n,i)}_{ab}\overline{h^{(TT,n,i)}_{a'b'}(x')}\,,\\
\delta^{(V)}_{aba'b'}(x,x') & = &
\sum_{n=2}^{\infty}
\sum_{i}V^{(n,i)}_{ab}(x)\overline{V_{a'b'}^{(n,i)}(x')}\,, \\ 
\delta^{(S)}_{aba'b'}(x,x') & = & 
\sum_{n=2}^{\infty}\sum_{i}W_{ab}^{(n,i)}(x)
\overline{W_{a'b'}^{(n,i)}(x')} 
 + \sum_{n=0}^{\infty}\sum_{i}\chi_{ab}^{(n,i)}
\overline{\chi_{a'b'}^{(n,i)}(x')}\,.
\end{eqnarray}
The Green function $G_{aba'b'}(x,x')$
of the operator $-{L_{ab}}^{cd}$ given by (\ref{Eulag1})
is defined by the equation
\begin{equation}
-L_x^{abcd} G_{cda'b'}(x,x') = {\delta^{ab}}_{a'b'}(x,x')\,.  \label{Gdel}
\end{equation}
Here, the subscript $x$ in $L_x^{abcd}$ 
indicates that the differential operator
$L^{abcd}$ acts at $x$ rather than at $x'$.

It can readily be seen that the Green function can be written in the form
\begin{equation}
G_{aba'b'}(x,x') = G^{(TT)}_{aba'b'}(x,x')
+G^{(V)}_{aba'b'}(x,x')+G^{(S)}_{aba'b'}(x,x')\,,
\end{equation}
with
\begin{eqnarray}
 G^{(TT)}_{aba'b'}(x,x') & = &
\sum_{n=2}^{\infty}\sum_{i} a_n
h^{(TT,n,i)}_{ab}\overline{h^{(TT,n,i)}_{a'b'}(x')}\,,\\
 G^{(V)}_{aba'b'}(x,x')  & = & 
\sum_{n=2}^{\infty}
\sum_{i} b_n V^{(n,i)}_{ab}(x)\overline{V_{a'b'}^{(n,i)}(x')}\,, \\ 
 G^{(S)}_{aba'b'}(x,x') & = &
\sum_{n=0}^{\infty}\sum_{i}c_n\chi_{ab}^{(n,i)}
\overline{\chi_{a'b'}^{(n,i)}(x')}
+ \sum_{n=2}^{\infty}\sum_{i} e_n W_{ab}^{(n,i)}(x)
\overline{W_{a'b'}^{(n,i)}(x')} \nonumber \\
&& + \sum_{n=2}^{\infty}\sum_{i} d_n \left[ W_{ab}^{(n,i)}(x)
\overline{\chi_{a'b'}^{(n,i)}(x')} 
+ \chi_{ab}^{(n,i)}(x)
\overline{W_{a'b'}^{(n,i)}(x')} \right]\,, \label{coeffs}
\end{eqnarray}
where $a_n$,$b_n$, $c_n$, $d_n$ and 
$e_n$ are constants.  We call the bi-tensors
$G^{(TT)}_{aba'b'}$ and $G^{(V)}_{aba'b'}$ the transverse-traceless and
vector sectors, respectively, and the $G^{(S)}_{aba'b'}$
the scalar sector of the propagator. It is clear that equation (\ref{Gdel})
is satisfied if
\begin{eqnarray}
-{L_{ab}}^{cd}G^{(TT)}_{cda'b'}(x,x')&=&\delta^{(TT)}_{aba'b'}(x,x')\,, 
\label{tensor} \\
-{L_{ab}}^{cd}G^{(V)}_{cda'b'}(x,x')&=&\delta^{(V)}_{aba'b'}(x,x')\,,
\label{vector} \\
-{L_{ab}}^{cd}G^{(S)}_{cda'b'}(x,x')&=&\delta^{(S)}_{aba'b'}(x,x')
\,. \label{scalar} 
\end{eqnarray}
We discuss the solutions of these equations in the next two sections.

\section{The transverse-traceless and vector sectors}

By applying the operator (\ref{Eulag1}) 
on the modes $h_{ab}^{(TT,n,i)}$ and $V_{ab}^{(n,i)}$ in the
transverse-traceless and vector sectors we find 
\begin{eqnarray}
-{L_{ab}}^{cd}h^{(TT,n,i)}_{cd} & = & 
\frac{\lambda_n}{2}h_{ab}^{(TT,n,i)}\,, \label{eqntensor} \\
-{L_{ab}}^{cd}V_{cd}^{(n,i)} & = & 
\frac{1}{2\alpha}\left(\lambda_n - 4H^2\right)V_{ab}^{(n,i)}\,.
\label{eqnvector}
\end{eqnarray}
One finds equation (\ref{eqnvector}) easily by noting that
the part of equation (\ref{Eulag1}) 
coming from the Lagrangian density ${\cal L}_{\rm inv}$ vanishes
because 
$V_{ab}^{(n,i)} \propto \nabla_a \xi^{(n,i)}_b + \nabla_b\xi^{(n,i)}_a$.
By using (\ref{eqntensor}) and (\ref{eqnvector}) in (\ref{tensor}) and 
(\ref{vector}) we find $a_n = 2\lambda_n^{-1}$ and 
$b_n =  2\alpha[\lambda_n - 4H^2]^{-1}$.  Hence, 
\begin{eqnarray}
G^{(TT)}_{aba'b'}(x,x') & =& 2\sum_{n=2}^{\infty}\sum_{j}
\frac{h^{(TT,n,j)}_{ab}(x)\overline{h^{(TT,n,j)}_{a'b'}(x')}}{\lambda_n}\,,\\
G^{(V)}_{aba'b'}(x,x') & = & 2\alpha \sum_{n=2}^{\infty}\sum_{j}
\frac{V^{(n,j)}_{ab}(x)\overline{V^{(n,j)}_{a'b'}(x')}}
{\lambda_n - 4H^2}\,.
\end{eqnarray}

Closed-form expressions for $G^{(TT)}_{aba'b'}(x,x')$ and 
$G^{(V)}_{aba'b'}(x,x')$ (with $\alpha=1$) have been derived by Allen and
Turyn~\cite{AllenTuryn}. We need to quote some definitions given in
\cite{AllenJacobson} to state their results.  We define
$\mu(x,x')$ to be the geodesic distance on $S^4$ between the two points
$x$ and $x'$.  Then, the vectors
$n_a \equiv \nabla_a \mu$, where the derivative operator
acts at $x$, and $n_{a'} \equiv \nabla_{a'}\mu$, where it 
acts at $x'$, are the tangent vectors to the geodesic.  (Primed indices
refer to the tangent space at $x'$ and unprimed indices refer to that at
$x$ here and in the rest of this paper.) 
Primed indices are raised and lowered by the metric $g_{a'b'}$ at
$x'$ and unprimed ones by the metric $g_{ab}$ at $x$.  The parallel
propagator ${g^a}_{a'}$ is defined as follows: given a vector $Y^{a'}$ at $x'$,
the vector ${g^a}_{a'}Y^{a'}$ is obtained by parallelly transporting $Y^{a'}$
along the geodesic to the point $x$.  We also define the variable
$z = \cos^2 (H\mu/2)$.  Large spacelike separation in de\ Sitter spacetime
corresponds
to the limit
$z\to -\infty$~\cite{AllenJacobson}. (Although 
there is no spacelike
geodesic connecting the two points for $z <-1$, large spacelike 
{\it coordinate} 
distance in the metric $ds^2 = (H\lambda)^{-2}(-d\lambda^2 + d{\bf x}^2)$
corresponds to this limit.)
We define further the following traceless bi-tensors:
\begin{eqnarray}
T^{(1)}_{aba'b'} & = & \left(n_a n_b-\frac{1}{4}g_{ab}\right)
\left(n_{a'} n_{b'}-\frac{1}{4}g_{a'b'}\right)\,,\label{T1} \\
T^{(2)}_{aba'b'} & = & g_{aa'}g_{bb'}+g_{a'b}g_{b'a}-\frac{1}{2}g_{ab}g_{a'b'}
\,,\label{T2} \\
T^{(3)}_{aba'b'} & = & g_{aa'}n_bn_{b'}+g_{ab'}n_{b}n_{a'}+g_{ba'}n_{a}n_{b'}
+g_{bb'}n_{a}n_{a'}+4n_{a}n_{b}n_{a'}n_{b'}\,. \label{T3}
\end{eqnarray}
With these definitions the bi-tensor
$G^{(TV)}_{aba'b'} \equiv G^{(TT)}_{aba'b'}(x,x') + G^{(V)}_{aba'b'}(x,x')$ 
can be written as 
\begin{equation}
 G^{(TV)}_{aba'b'}(x,x')=\frac{H^2}{16\pi^2}\left[f^{(TV,1)}(z)
T^{(1)}_{aba'b'}
+f^{(TV,2)}(z)T^{(2)}_{aba'b'}+f^{(TV,3)}(z)T^{(3)}_{aba'b'} \right]\,,
\end{equation}
where
\begin{eqnarray}
 f^{(TV,1)}(z)  & = & -\frac{16}{9} + \frac{8\alpha}{3}
+ \frac{4-4\alpha}{1-z}\nonumber \\
&& + \left( 1- \frac{3\alpha}{5}\right)
\left[ -\frac{8}{3}\left(2- \frac{1}{z^2} - \frac{1}{z^3}\right)
\log (1-z) +\frac{4}{z} + \frac{8}{3z^2} \right]\,,\\
 f^{(TV,2)}(z)  & = &\frac{2}{3}-\frac{13\alpha}{5} + \frac{1+3\alpha}{6(1-z)}
\nonumber \\
&& + \left(1-\frac{3\alpha}{5}\right)\left[
\left( 2 + \frac{1}{9z^2} + \frac{1}{9z^3}\right)\log (1-z) 
+\frac{1}{6z}+\frac{1}{9z^2}\right]\,, \\
 f^{(TV,3)}(z) 
 & = & \frac{6\alpha z}{5} + \frac{2}{3}-\frac{16\alpha}{5} -
 \frac{2}{3(1-z)}\nonumber \\
 && + \left(1-\frac{3\alpha}{5}\right)
\left[\left( 2 - \frac{10}{9z} - \frac{4}{9z^2} - \frac{4}{9z^3}\right)
\log{(1-z)} 
-\frac{2}{3z}- \frac{4}{9z^2}\right]\,.
\end{eqnarray}
(The $G^{(TV)}_{aba'b'}$ here is twice as large as that of Allen
and Turyn due to the difference in the normalization of the field $h_{ab}$.)
It is interesting to note that logarithmic terms are absent 
in $G^{(TV)}_{aba'b'}(x,x')$ if we choose
$\alpha = 5/3$.  
Notice, however, that there is a term linear in $z$ unless $\alpha=0$.

\section{The scalar sector}

Application of the operator defined by (\ref{Eulag1}) on the tensor 
eigenfunctions 
appearing in the scalar sector yields, after a long but straightforward
calculation,
\begin{eqnarray}
-L_{ab}}^{cd}{\chi_{cd}^{(n,i)} & = &  
K^{(n)}_{11}\chi^{(n,i)}_{ab} +K_{12}^{(n)}W^{(n,i)}_{ab}\,,  \label{mix1} \\
-{L_{ab}}^{cd}W_{cd}^{(n,i)}  
& =& 
K_{12}^{(n)}\chi_{ab}^{(n,i)} + K_{22}^{(n)}W_{ab}^{(n,i)}\,, \label{mix2}
\end{eqnarray}
where 
\begin{eqnarray}
K^{(n)}_{11} & = &\left[\frac{1}{\alpha}\left(\frac{3}{2}
+\frac{2}{\beta}\right)^2-\frac{3}{4}\right]\lambda_{n}+3H^2\,,\\
K^{(n)}_{12}  & = & -\left(\frac{\alpha-3}{4\alpha}-\frac{1}{\alpha \beta}
\right)\sqrt{3\lambda_{n}(\lambda_{n}-4H^2)}\,,\\
K^{(n)}_{22} & = & -\frac{\alpha-3}{4\alpha}
\left(\lambda_{n}+\frac{12H^2}{\alpha-3} \right)\,.
\end{eqnarray}
The equation 
$-{L_{ab}}^{cd}G_{cda'b'}^{(S)}(x,x')=\delta_{aba'b'}^{(S)}(x,x')$
is satisfied if
\begin{equation}
\left ( \begin{array}{cc} c_{n} & d_{n} \\ d_{n} & e_{n} \end{array}\right )
= \left (\begin{array}{cc} K^{(n)}_{11} & K^{(n)}_{12} \\ 
K^{(n)}_{12} & K^{(n)}_{22} \end{array} \right)^{-1}\,.
\end{equation}
One can readily find the $G^{(S)}_{aba'b'}(x,x')$ by 
substituting the coefficients $c_n$, $d_n$ and $e_n$ thus obtained\footnote{The
cases with $n=0$ and $1$ need to be treated separately, but the expressions
for $c_n$ for these cases turn out to be the same as the other cases.} in
(\ref{coeffs}) and recalling the definitions 
of $\chi_{ab}^{(n,i)}$ and $W_{ab}^{(n,i)}$ given by (\ref{defchi}) and 
(\ref{defW}).  Let us define
\begin{eqnarray}
\Delta_{m^2}(x,x') & = & 
\sum_{n=0}^\infty \sum_i \frac{\phi^{(n,i)}(x)\overline{\phi^{(n,i)}(x')}}
{\lambda_n + m^2 H^2}\,,\\
\Delta^{-}_{m^2}(x,x') & = & 
\sum_{n=1}^\infty \sum_i \frac{\phi^{(n,i)}(x)
\overline{\phi^{(n,i)}(x')}}{\lambda_n+m^2 H^2}\,,\\
\Delta^{- -}_{m^2}(x,x') & = & 
\sum_{n=2}^\infty \sum_i \frac{\phi^{(n,i)}(x)\overline{\phi^{(n,i)}(x')}}
{\lambda_n + m^2 H^2}\,,\\
\Delta^{(1)}_{m^2}(x,x') & = & 
\sum_{n=0}^\infty \sum_i \frac{\phi^{(n,i)}(x)\overline{\phi^{(n,i)}(x')}}
{(\lambda_n + m^2 H^2)^2}\,.
\end{eqnarray}
Then, we can write
\begin{eqnarray}
 G^{(S)}_{aba'b'}(x,x') & = &
\left( \nabla_a \nabla_b-\frac{1}{4}g_{ab}\Box\right)
\left( \nabla_{a'} \nabla_{b'}-\frac{1}{4}g_{a'b'}\Box'\right)
A^{(1)}_{n}(x,x')
\nonumber \\
&& +\frac{1}{2}\left[\left( \nabla_a \nabla_b-\frac{1}{4}g_{ab}\Box\right)
g_{a'b'}+g_{ab}\left( \nabla_{a'} \nabla_{b'}
-\frac{1}{4}g_{a'b'}\Box'\right)\right]A^{(2)}(x,x')
\nonumber \\
&& +\frac{1}{4}g_{ab}g_{a'b'} A^{(3)}_{n}(x,x')\,,
\label{differentiation}
\end{eqnarray}
where
\begin{eqnarray}
A^{(1)} & = & \frac{\alpha}{9H^4}\Delta^{-}_0
- \frac{1}{3H^4}\Delta^{- -}_{-4}
+ \frac{3-\alpha}{9H^4}\Delta_{3\beta} + 
\frac{4-(\alpha-3)\beta}{3H^2}\Delta^{(1)}_{3\beta}\,, \\
A^{(2)} & = & 
\frac{\beta[4 - (\alpha-3)\beta]}{2}\Delta^{(1)}_{3\beta}\,,\\
A^{(3)} & = & \frac{\beta^2}{4}\left\{ (\alpha-3)\Delta_{3\beta}
+ 3[4-(\alpha-3)\beta]H^2\Delta^{(1)}_{3\beta}\right\}\,.
\end{eqnarray}
(The operators $\Box$ and $\Box'$ act at $x$ and $x'$, respectively.)
We have used the fact that the differential operator 
$\nabla_a \nabla_b - \frac{1}{4}g_{ab}\Box$ annihilates the modes
$\phi^{(n,i)}(x)$ with $n=0$ or $1$.

One can explicitly evaluate $G^{(S)}_{aba'b'}(x,x')$ in terms of the variable 
$z = \cos^2(H\mu/2)$ by using~\cite{AllenJacobson}
\begin{equation}
\Delta_{m^2}  =  \frac{H^2}{16\pi^2}\Gamma(a_+)\Gamma(a_-)
F(a_+,a_-;2;z)\,, \label{hyper}
\end{equation}
where 
\begin{equation}
a_{\pm} = \frac{3}{2}\pm \left( \frac{9}{4} - m^2 \right)^{1/2}\,,
\end{equation}
and~\cite{AllenTuryn}
\begin{eqnarray}
\Delta_{m^2}^{(1)}  & = &
- H^{-2} \frac{\partial\ }{\partial c} \Delta_{c}(x,x')|_{c=m^2}\,, \\
\Delta^{-}_0 & = & 
\frac{H^2}{16\pi^2}\left[\frac{1}{1-z} - 2\log(1-z) - \frac{14}{3}
\right]\,, \\
\Delta^{- -}_{-4}  & = &
\frac{H^2}{16\pi^2}
\left[\frac{1}{1-z}+6(1-2z)\log(1-z)
+ \frac{1}{10}(67-224z)\right]\,.
\end{eqnarray}
Here $\Gamma(x)$ is the gamma function and $F(a,b;c;x)$ is Gauss' 
hypergeometric function.
The modulus of the function $\Delta_{m^2}$ behaves like $|z|^{-3/2}$ if
$m^2 \geq 3/2$ and like $|z|^{-a_-}$ if $m^2 < 3/2$ for large $|z|$
provided that $a_{-}$ is not zero or a negative integer.
The gauge used by Allen and Turyn corresponds to the choice $\alpha=1$,
$\beta = -2$.  Therefore their propagator involves the function
$\Delta_{-6}$ which increases like $|z|^{(\sqrt{33}-3)/2}$ at large $|z|$.
One can make the scalar sector $G^{(S)}_{aba'b'}$ decrease for large $|z|$
by choosing $\beta > 0$~\cite{HigKou2}. The choice $\beta = 2/3$ is
particularly convenient since the functions $\Delta_{3\beta}$ and
$\Delta^{(1)}_{3\beta}$ in this case are 
\begin{eqnarray}
\Delta_2 & = & \frac{H^2}{16\pi^2}\frac{1}{1-z}\,, \\
\Delta_2^{(1)}
& = & - \frac{1}{16\pi^2}\frac{\log(1-z)}{z}\,.
\end{eqnarray}
We will write down $G^{(S)}_{aba'b'}$ as a function of $z$ explicitly with
this choice of $\beta$.
For this purpose the following formulas are useful~\cite{AllenJacobson}:
\begin{eqnarray}
\nabla_{a}n_{b}  & = &  H\cot (\mu H)(g_{ab}-n_a n_b) =
\frac{H(2z-1)}{2\sqrt{z(1-z)}}(g_{ab}-n_a n_b)\,,\\
\nabla_{a}n_{b'}  & = & -\frac{H}{\sin(\mu H)}(g_{ab'}+n_a n_{b'})
= - \frac{H}{2\sqrt{z(1-z)}}(g_{ab'}+n_a n_{b'})\,,\\
\nabla_{a}g_{bb'} & = &
\frac{H[1-\cos(\mu H)]}{\sin(\mu H)}(g_{ab}n_{b'}+g_{ab'} n_b)
= H\sqrt{\frac{1-z}{z}}(g_{ab}n_{b'}+g_{ab'} n_b)\,.
\end{eqnarray}
By using these formulae one finds~\cite{AllenTuryn} 
\begin{equation}
\left( \nabla_a \nabla_b - \frac{1}{4}g_{ab}\Box\right)\Phi(z)
= H^2 z(1-z)\frac{d^2\Phi}{dz^2} \left( n_a n_b - \frac{1}{4}g_{ab}\right)\,.
\end{equation}
We also use
\begin{eqnarray}
&& \left(\nabla_{a'}\nabla_{b'} - \frac{1}{4}g_{a'b'}\Box'\right)
\left(\nabla_{a}\nabla_{b} - \frac{1}{4}g_{ab}\Box \right)\Phi(z) \nonumber \\
&& = H^4 \left\{ z(1-z)\frac{d^2\ }{dz^2}
\left[ z(1-z)\frac{d^2\Phi}{dz^2}\right]T^{(1)}_{aba'b'}\right. \nonumber \\
&& \ \ \ + \left. \frac{1}{4}\frac{d^2\Phi}{dz^2}\,T^{(2)}_{aba'b'}
+ \frac{1-z}{2}\frac{d\ }{dz}\left[ z \frac{d^2\Phi}{dz^2}\right]
\,T^{(3)}_{aba'b'}\right\} \,.
\end{eqnarray}
The result is then given in the following form:
\begin{equation}
G^{(S)}_{aba'b'}(x,x')|_{\beta=2/3}
= \frac{H^2}{16\pi^2}\sum_{k=1}^5 f^{(S,k)}(z)T^{(k)}_{aba'b'}\,,
\end{equation}
where $T^{(4)}_{aba'b'}$ and $T^{(5)}_{aba'b'}$ are defined as
\begin{eqnarray}
T^{(4)}_{aba'b'} & = & g_{ab}\left(n_{a'}n_{b'}-\frac{1}{4}g_{a'b'}\right)+
g_{a'b'}\left(n_{a}n_{b}-\frac{1}{4}g_{ab}\right)\,,\\
T^{(5)}_{aba'b'} & = & g_{ab}g_{a'b'}\,.
\end{eqnarray}
The functions $f^{(S,k)}(z)$ are
\begin{eqnarray}
f^{(S,1)}(z) & = & \frac{4(\alpha - 9)}{9}\left[ \left(
\frac{2}{z}-\frac{8}{z^2}
+ \frac{6}{z^3}\right)\log (1-z)
- \frac{5}{z}+\frac{6}{z^2}\right]\,, \\
f^{(S,2)}(z)  & = & 
\left( \frac{\alpha}{18} - \frac{3}{2}\right)\frac{1}{1-z}
+ \frac{\alpha-9}{18}\left[
\frac{2}{z^3}\log(1-z)
+ \frac{1}{z} + \frac{2}{z^2}\right]\,, \\
f^{(S,3)}(z) & = & \left( \frac{\alpha}{9} -3\right)\frac{1}{1-z}
+ \frac{2(\alpha-9)}{9}\left[
\left(\frac{2}{z^2}-\frac{2}{z^3}\right)\log(1-z)
+ \frac{1}{z} - \frac{2}{z^2}\right]\,, \\
f^{(S,4)}(z) & = & \left( 1-\frac{\alpha}{9}\right)
\left[ \frac{1}{1-z}- \frac{2}{z}
+ \left(\frac{2}{z}-\frac{2}{z^2}\right)\log(1-z)\right]\,,\\
f^{(S,5)}(z) & = & \left( \frac{\alpha}{36}-\frac{1}{12}\right)\frac{1}{1-z}
+ \left(\frac{\alpha}{18}-\frac{1}{2}\right)\frac{\log (1-z)}{z}\,.
\end{eqnarray}
The $G^{(S)}_{aba'b'}$ simplifies considerably if we let $\alpha=9$.

\section{Summary and Discussions}

The full propagator can be obtained by adding the transverse-traceless,
vector and scalar sectors as
\begin{equation}
G_{aba'b'}(x,x')|_{\beta=2/3} = \frac{H^2}{16\pi^2}
\sum_{k=1}^5 [f^{(TV,k)}(z) + f^{(S,k)}(z)]T^{(k)}_{aba'b'} \label{propa}
\end{equation}
with the definition $f^{(TV,4)}(z) = f^{(TV,5)}(z) = 0$.

Let us first comment on the relation between the results here
and our previous work~\cite{HigKou2}
which concentrated on the scalar sector.  In \cite{HigKou2}
we did not adopt the Euclidean approach, and therefore the definition of
the scalar sector was slightly different.  
Let us assume $1-z \neq 0$ so that the two points are not on the light cone of
one another.
Then we have 
\begin{eqnarray}
(\Box - m^2 H^2)\Delta_{m^2}  &= & 0\,,\nonumber \\
(\Box - m^2 H^2)\Delta^{(1)}_{m^2} & = & -\Delta_{m^2}\,, \nonumber\\
\Box \Delta^{-}_0 & = & \frac{3H^4}{8\pi^2}\,.\nonumber
\end{eqnarray}
By using these formulas one can write the
scalar sector as 
\begin{equation}
G^{(S)}_{aba'b'}(x,x') = G^{(S1)}_{aba'b'}(x,x') + G^{(S2)}_{aba'b'}(x,x')\,,
\end{equation}
where
\begin{eqnarray}
G^{(S1)}_{aba'b'} & = & 
\nabla_a \nabla_b \nabla_{a'}\nabla_{b'}
\left( \frac{3-\alpha}{9H^4}\Delta_{3\beta}
+ \frac{4-(\alpha-3)\beta}{3H^2}\Delta^{(1)}_{3\beta}\right) \nonumber \\
&& + \left( g_{ab}\nabla_{a'}\nabla_{b'} + g_{a'b'}\nabla_a \nabla_b\right)
\frac{1}{3H^2}\Delta_{3\beta}\,, \\
G^{(S2)}_{aba'b'} & = &  \frac{\alpha}{9H^4}
\nabla_a \nabla_b \nabla_{a'}\nabla_{b'}\Delta^{-}_0 \nonumber \\
&& - \frac{1}{3H^4}\left( \nabla_a \nabla_b-\frac{1}{4}g_{ab}\Box\right)
\left( \nabla_{a'} \nabla_{b'}-\frac{1}{4}g_{a'b'}\Box'\right)\Delta^{- -}_{-4}
\,.
\end{eqnarray}
The $G^{(S1)}_{aba'b'}$ is identical with the scalar sector of
\cite{HigKou2} for $\beta = 4/(\alpha-3)$.
The $G^{(S2)}_{aba'b'}$ clearly satisfies
$g^{ab}G^{(S2)}_{aba'b'} = 0$ and can also be shown to satisfy 
$\nabla^a\nabla^b G^{(S2)}_{aba'b'} = 0$.  Therefore, the non-scalar sector
as defined in \cite{HigKou2} is 
$G^{(TV)}_{aba'b'} + G^{(S2)}_{aba'b'}$.
It is also clear that the terms that depend on the gauge parameters are
of pure-gauge form, i.e.
$\nabla_{(a} W^{(1)}_{b)a'b'} + \nabla_{(a'} W^{(2)}_{b') ab}$
for some $W^{(1)}_{ba'b'}$ and $W^{(2)}_{b'ab}$, where $(\cdot\cdot\cdot)$ 
denotes
symmetrization.  As expected,
the two-point function of a gauge-invariant quantity does not depend on
the gauge parameters 
because the pure-gauge part of the propagator will
not contribute. 

Finally, let us show that the 
two-point function of a gauge-invariant
quantity obtained by applying differential operators on the field $h_{ab}$,
which we call a local gauge-invariant quantity, 
is bounded as $z \to -\infty$ even though the propagator given by
(\ref{propa}) is not bounded for any $\alpha$.
(One example of a local gauge-invariant quantity 
is the linearized Weyl tensor.)
By choosing $\alpha=0$ the linear term in $G_{aba'b'}$ is eliminated,
and its growth for large $|z|$ is logarithmic.
We will see that this growth is not reflected in the two-point function of a
local gauge-invariant quantity.

First we note that the bi-vector
\begin{equation}
A_{aa'} = \phi_1(z) g_{aa'} + \phi_2(z) n_a n_{a'}
\end{equation}
is divergence-free if
\begin{eqnarray}
\phi_1(z) & = & -\frac{2}{3}z(1-z)f(z)+ (2z-1)\int^z f(z)dz\,, \\
\phi_2(z) & = & -\frac{2}{3}z(1-z)f(z)+ 2(z-1)\int^z f(z)dz
\end{eqnarray}
for some function $f(z)$~\cite{AllenTuryn}.  Next we find
\begin{eqnarray}
&& \nabla_a \nabla_{a'}A_{bb'} + 
\nabla_b \nabla_{a'}A_{ab'} + 
\nabla_a \nabla_{b'}A_{ba'} + 
\nabla_b \nabla_{b'}A_{aa'}  \nonumber \\
&& = F^{(1)}(z)T^{(1)}_{aba'b'}
+ F^{(2)}(z)T^{(2)}_{aba'b'}
+ F^{(3)}(z)T^{(3)}_{aba'b'}\,, \label{useful}
\end{eqnarray}
where
\begin{eqnarray}
F^{(1)}(z)  & = & -\frac{16}{3}z(1-z)f'(z)\,, \\
F^{(2)}(z)  & = & \frac{4}{3}(2z-1)f(z) - \frac{2}{3}z(1-z)f'(z)\,, \\
F^{(3)}(z)  & = & - \frac{2}{3}z^2(1-z)^2f''(z) + \frac{4}{3}(3z-2)z(1-z)f'(z)
- \frac{8}{3}(1-z)^2f(z)\,.
\end{eqnarray} 
(The left-hand side of (\ref{useful}) is guaranteed to be traceless and, 
therefore, is a linear combination of $T^{(1)}_{aba'b'}$, $T^{(2)}_{aba'b'}$
and $T^{(3)}_{aba'b'}$, because
$A_{aa'}$ is divergence-free.)
By using (\ref{useful}) with
$f(z) = 1/(1-z)+1/[2(1-z)^2]$ we find that the bi-tensor
\begin{equation}
P_{aba'b'} \equiv
\left[\frac{16}{3} -\frac{16}{3(1-z)^2}\right]T^{(1)}_{aba'b'}
- 2T^{(2)}_{aba'b'} + \left[ -2 - \frac{4}{3(1-z)} - \frac{2}{3(1-z)^2}\right]
T^{(3)}_{aba'b'}
\end{equation}
is of pure-gauge form. Notice that for large $|z|$ we have (for $\beta > 0$)
\begin{equation}
G_{aba'b'}|_{\alpha=0} \approx -\frac{H^2}{16\pi^2}\log(1-z)\,P_{aba'b'}\,,
\end{equation}
where $P_{aba'b'}$ is bounded as $z\to -\infty$.
Now, suppose that we want to find the two-point function of a local 
gauge-invariant quantity at $z=z_0$.  We can use the propagator
\[
G_{aba'b'}(x,x')|_{\alpha=0} + \frac{H^2}{16\pi^2}\log(1-z_0)\,P_{aba'b'}
(x,x')
\]
with $z_0$ fixed
for this calculation because $P_{aba'b'}$ is of pure-gauge form.
This propagator and its derivatives at $z=z_0$
are bounded functions of $z_0$ as $|z_0|$ becomes
large.  This implies that
the logarithmic growth of the propagator will not be
reflected in the two-point function of a local gauge-invariant quantity,
and that the latter is bounded as $z\to -\infty$.

Although this argument shows that two-point functions of local gauge-invariant
quantities do not
grow at large distances, it does not show whether or not they decrease, or
if so, how rapidly.  One needs to compute such quantities explicitly to
find their detailed
large-distance behaviour.  One of the authors (SSK) is currently investigating
the two-point function of the linearized Weyl tensor.  The result of this
calculation will be reported elsewhere~\cite{Spyros}.

\

\begin{flushleft}
\large{\bf Acknowledgements}
\end{flushleft}

We thank John Friedman, Bernard Kay, Bill Unruh and Bob Wald
for useful discussions.  We used Maple 6 for some of our calculations.

\

\end{document}